\documentclass[prl,twocolumn,superscriptaddress,showpacs,amsmath,amssymb]{revtex4}
\usepackage{graphicx}
\usepackage{bm}
\usepackage{amsmath} 
\usepackage{subfigure}

\newcommand{\lindblad}{\mathcal{D}}

\newcommand{\hsys}{H_\mathrm{sys}}
\newcommand{\hdrive}{H_\mathrm{drive}}
\newcommand{\hsb}{H_\mathrm{sb}}
\newcommand{\hphon}{H_\mathrm{env}}
\newcommand{\hmeas}{H_\mathrm{meas}}
\newcommand{\hleads}{H_\mathrm{leads}}

\newcommand{\source}{S}
\newcommand{\drain}{D}
\newcommand{\hc}{\mathsf{H.c.}}
\newcommand{\bra}[1]{\langle#1\vert}
\newcommand{\ket}[1]{\vert#1\rangle}
\newcommand{\calT}{\mathcal{T}}
\newcommand{\calL}{\mathcal{L}}
\DeclareMathOperator{\Tr}{Tr}
\newcommand{\commute}[2]{[#1,#2]}
\DeclareMathOperator{\ramp}{\Theta}
\newcommand{\eqn}[1]{Eq.~(\ref{#1})}

\begin{document}

\title{Continuous measurement of a microwave-driven solid state qubit}


\author{S. D. Barrett}
\email{sean.barrett@hp.com}
 \affiliation{Hewlett-Packard
Laboratories, Filton Road, Stoke Gifford, Bristol BS34 8QZ, U.K.}

\author{T. M. Stace}
\email{tms29@cam.ac.uk} \affiliation{DAMTP, University of
Cambridge, Wilberforce Road, CB3 0WA, U.K.}

\date{\today}


\begin{abstract}
We analyze the dynamics of a continuously observed, damped,
microwave driven solid state charge qubit. The qubit consists of a
single electron in a double well potential, coupled to an
oscillating electric field, and which is continuously observed by
a nearby point contact electrometer. The microwave field induces
transitions between the qubit eigenstates, which have a profound
effect on the detector output current. We show that useful
information about the qubit dynamics, such as dephasing and
relaxation rates, and the Rabi frequency, can be extracted from
the DC detector conductance and the detector output noise power
spectrum. We also demonstrate that these phenomena can be used for
single shot electron \emph{spin} readout, for spin based quantum
information processing.
\end{abstract}
\maketitle

Recently, rapid experimental progress in mesoscopic physics has
meant that it is now possible to confine, manipulate, and measure
small numbers of electrons in single or coupled quantum dots
\cite{ManyExperiments,hay03,Elzerman2004,pet04}. The mesoscopic
environment of such confined electron systems may consist of
phonons, electrons, and other electromagnetic degrees of freedom.
Thus these experiments are particularly interesting as they allow
the complex interaction between such confined systems and their
environment to be studied at the single electron level.
Furthermore, in view of the potential applications of such systems
in solid state quantum information processing \cite{los98},
understanding these interactions is important, as it allows qubit
decoherence mechanisms to be studied and accurately characterised.

Of particular interest are coupled quantum dot (CQD) qubit systems
driven by oscillating electric fields. The presence of a driving
field resonant to the qubit energy splitting can drive transitions
into the excited state. Continuous measurement of such systems can
reveal important spectroscopic information about the qubit, such
as the qubit splitting, Rabi frequency, and decoherence rates.
Earlier work has focussed on current transport through driven,
open CQD systems \cite{Oosterkamp98,Brandes2004}. In a recent
experiment, a driven qubit comprising a single electron in a
closed CQD system was continuously observed via a nearby quantum
point contact (QPC) detector \cite{pet04}.

\begin{figure}[t]
\includegraphics[width=7cm]{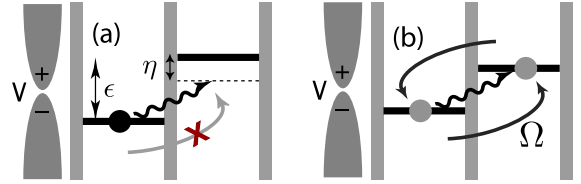}
\caption{Schematic of a CQD charge qubit measured by a QPC under
(a) non-resonant and (b) resonant microwave driving.}
\label{fig:cartoon}
\end{figure}

In this Letter, we theoretically analyse this system (see Fig.
\ref{fig:cartoon}). We account for the coupling of the CQD both to
the QPC and to a generic bosonic environment, which may comprise
of phonons or other electromagnetic degrees of freedom
\cite{BarrettMilburn2003}. Both the detector and the environment
contribute to the qubit relaxation and dephasing rates. We also
demonstrate how resonant driving phenomena can be used for
single-shot readout of the electron spin. Our results are also
relevant to other driven qubit systems, e.g. superconducting
charge qubits \cite{YaleDrivenQubit,Duty04}.

A number of authors have considered the continuous measurement of
\emph{undriven} charge qubit systems by a QPC detector
\cite{ManyTheoryQPC,Goan2001,Goan2001b}. Continuous measurement of
driven superconducting flux qubits has also been considered
\cite{Smirnov03}. In what follows, we adopt the quantum
trajectories description of the measurement process
\cite{gar00,Goan2001,Goan2001b}, and generalize results obtained
in previous work \cite{stace:136802,sta03c} on the measurement of
\emph{undriven} charge qubits using  a QPC at arbitrary bias
voltage. We first derive a master equation (ME) for the dynamics
of the damped, driven charge qubit system. We use solutions of the
ME to determine the DC conductance and current power spectra of
the QPC detector, and show how various qubit parameters can be
extracted from measurements of these quantities. Finally, we
describe our technique for spin readout.

The model system  we consider is shown in Fig. \ref{fig:cartoon},
for which the total Hamiltonian is given by
$H=\hsys+\hdrive+\hmeas+\hleads+\hsb+\hphon $, where ($\hbar = 1$)
\begin{align}
\hsys &= (-\Delta\sigma_x-\epsilon\sigma_z)/2 \equiv -\phi\sigma_z^{(e)}/2,\label{eqn:hsys}\\
\hdrive & = \Omega_0 \cos[(\phi-\eta)t] \left(\cos\delta
\,\sigma_z
+ \sin\delta \, \sigma_x \right) ,\\
\hmeas&=\sum_{k,q} \lambda (\calT + \nu \sigma_z)
a^\dagger_{\drain,q} a_{\source,k}+\hc , \label{eqn:hmeas} \\
%
%
\hleads&=\sum_k \omega_{\source,k} a^\dagger_{\source,k}
a_{\source,k} + \sum_q \omega_{\drain,q} a^\dagger_{\drain,q} a_{\drain,q}, \\
\hsb & = \sigma_z \sum_i \kappa_i (b_i^\dag + b_i) , \quad
\hphon  = \sum_i \omega_{B,i} b_i^\dag b_i . \label{eqn:hphon}
\end{align}
Here, $\hsys$ is the bare qubit Hamiltonian, in which $\sigma_x =
\ket{l}\bra{r}+\ket{r}\bra{l}$ and $\sigma_z =
\ket{l}\bra{l}-\ket{r}\bra{r}$, where $\ket{l}$ ($\ket{r}$)
denotes an electron state localized on the left (right) dot, $\phi
= \sqrt{\Delta^2+\epsilon^2}$ is the qubit energy splitting, and
$\sigma_z^{(e)}=\ket{g}\bra{g}-\ket{e}\bra{e}$. $\hdrive$
corresponds to the driving field with frequency $\omega_0 = \phi -
\eta$, which may couple to both the $\sigma_x$ and $\sigma_z$
qubit operators, as parameterized by $\delta$. $\hmeas$ denotes
the qubit-detector coupling, in terms of the dimensionless
tunnelling parameters $\calT = \sqrt{2\pi g_\source g_\drain} T$
and $\nu = \sqrt{2\pi g_\source g_\drain} \chi$, and $\lambda =
1/\sqrt{2\pi g_\source g_\drain}$. We have assumed that the
tunnelling amplitudes, $T$ and $\chi$, and the densities of lead
modes, $g_\source$ and $g_\drain$, are approximately independent
of $k$ and $q$ over the energy range where tunnelling is allowed.
$\hleads$ is the free Hamiltonian of the source and drain leads,
where $a_{\source,j}$ ($a_{\drain,j}$) is the annihilation
operator for an electron in the $j$th source (drain) mode. $\hsb$
and $\hphon$ correspond to a standard spin-boson coupling to a
generic bath of bosons \cite{Legget1987}, where $b_i$ is the
annihilation for the $i$th boson mode.

To simplify the Hamiltonian, we first transform to an interaction
picture defined by $H_0=-\omega_0
\sigma_z^{(e)}/2+\hleads+\hphon$, and make our first rotating wave
approximation (RWA) and neglect terms oscillating rapidly compared
to $\Omega_0$ and the microwave detuning, $\eta$. Under this
transformation, the time dependence is transferred to the
interaction term and the Hamiltonian becomes
\begin{equation}
H_I(t)=-\frac{\eta}{2}\sigma_z^{(e)}-\frac{\Omega}{2}
\sigma_x^{(e)}  + A_I(t) \otimes Y(t)  + B_I(t) \otimes Z(t),
\end{equation}
where $\Omega=\Omega_0 \sin(\theta-\delta)$. We have defined the
system operators $A_I(t) = \sum_n e^{-i \omega_n t} P_n$ and
$B_I(t) = \sum_n e^{-i \omega_n t} Q_n$, with $\omega_n = 0, \pm
\omega_0$, $P_1 = \calT + \nu \cos \theta \sigma_z^{(e)}$, $P_2 =
P_3^\dag = -\nu \sin \theta \ket{e}\bra{g}$, $Q_1 = \cos \theta
\sigma_z^{(e)}$, and $Q_2 = Q_3^\dag = - \sin \theta
\ket{e}\bra{g}$. We have also defined the operators $Y(t) =
\lambda \sum_{k,q}
(e^{-i(\omega_{\source,k}-\omega_{\drain,q})t}a^\dagger_{\drain,q}
a_{\source,k}+\hc)$ and $Z(t) = \sum_i \kappa_i (e^{-i
\omega_{B,i} t} b_i^\dag + \hc)$ which act on the QPC and bosonic
environment degrees of freedom, respectively.

To derive the ME for the dynamics of the qubit alone, we further
transform to a frame defined by
$H_0'=-\frac{\eta}{2}\sigma_z^{(e)}-\frac{\Omega}{2}
\sigma_x^{(e)}$, in which all the dynamics are contained in the
qubit-QPC and qubit-environment interaction terms.  Then
$H_{I'}(t)= A_{I'}(t) \otimes Y(t)  + B_{I'}(t) \otimes Z(t)$,
where $A_{I'}(t)=\sum_{nn'} e^{-i (\omega_n+\omega_n') t} P_{nn'}$
and $B_{I'}(t)=\sum_{nn'} e^{-i (\omega_n+\omega_n') t} Q_{nn'}$,
for some operators $P_{nn'}$ and $Q_{nn'}$, and $\omega_{n'}=0,\pm
\Omega'$ where $\Omega'=\sqrt{\Omega^2+\eta^2}$. In this picture,
the qubit density matrix, $\rho$, satisfies
\begin{equation}
  \dot \rho_{I'}(t)=-\Tr_{\source,\drain,B}
  \{
  \int_{0}^t dt' \commute{H_{I'}(t)}{ \commute{H_{I'}(t')}{R(t')} }
  \}.
\label{eqn:SecondOrderExpansion}
\end{equation}
We now make a Born-Markov approximation, setting the lower limit
of the integral to $-\infty$ and
$R(t')=\rho_{I'}(t)\otimes\rho_{\source}\otimes\rho_{\drain}\otimes\rho_{B}$,
where $\rho_{\source}$, $\rho_{\drain}$, and $\rho_{B}$ are
equilibrium density matrices for the source, drain, and bath
degrees of freedom. At this point, it is convenient to introduce
the asymmetric quantum noise power spectra
\cite{Aguado2000,Schoelkopf2002} for the QPC and bath,
$S_Y(\omega) = \int_{-\infty}^{\infty} dt e^{i \omega t}
\Tr[Y(t)Y(0)\rho_{\source}\otimes\rho_{\drain}] =
\Theta(\omega-eV)+\Theta(\omega+eV)$ and  $S_Z(\omega) =
\int_{-\infty}^{\infty} dt e^{i \omega t} \Tr[Z(t)Z(0)\rho_{B}] =
2 \pi J(\omega)[1+n(\omega)]+ 2 \pi J(-\omega)n(-\omega)$, where
$\ramp(x)=(x+|x|)/2$ is the ramp function, $e V$ is the
source-drain bias across the detector, $J(\omega) = \sum_i
\kappa_i^2 \delta(\omega-\omega_{B,i})$ is the bath spectral
density, and $n(\omega)$ is the thermal equilibrium Bose
occupation number. To proceed, we make a second RWA, where we
neglect terms in Eq.(\ref{eqn:SecondOrderExpansion}) rotating at a
rate $\omega_0$. This RWA is justified in the limit of weak
coupling to the detector and environment, $\omega_0 \gg
S_{Y,Z}(\omega_n)$ \footnote{This RWA implies a course graining in
time, which means that our treatment does not describe dynamics on
very short timescales of order $\phi^{-1}$.} . We also assume that
the driving field is sufficiently weak that
$S_{Y,Z}(\omega_n+\omega'_{n'}) \approx S_{Y,Z}(\omega_n)$, i.e.
that the noise spectra are slowly varying over frequencies of
order $\Omega'$. Finally, in order to treat dephasing within our
perturbative technique, we require that $\lim_{\omega \to 0}
J(\omega) \propto \omega^s$ where $s \ge 1$, i.e. that the bath is
ohmic or superohmic at low frequencies.

These approximations allow us to derive a ME which is valid for
arbitrary source-drain bias and arbitrary bath temperature.
However, in this Letter, we restrict our attention to the low bias
($eV < \phi$) and low temperature ($kT \ll \phi$) regime, which
has been probed in a recent \cite{pet04}. In this case, the master
equation for the qubit in the original interaction picture is
given by
\begin{align}
\dot \rho_I(t) =&
-i\commute{-\frac{\eta}{2}\sigma_z^{(e)}-\frac{\Omega}{2}\sigma_x^{(e)}}{\rho_I(t)}
\nonumber
\\
&+\frac{1}{2}\Gamma_{\varphi} \lindblad[\sigma_z^{(e)}]\rho_I(t)
+\Gamma_{r}   \lindblad[\ket{g}\bra{e}]\rho_I(t) \nonumber \\
\equiv &\calL_I \rho_I(t), \label{eqn:mastereqn2}
\end{align}
where $\Gamma_{\varphi} = \Gamma_{\varphi}^\mathrm{det} +
\Gamma_{\varphi}^\mathrm{env}$ is the pure dephasing rate, with
$\Gamma_{\varphi}^\mathrm{det}= 2\nu^2 \cos^2\theta S_{Y}(0)$ and
$\Gamma_{\varphi}^\mathrm{env}= 2 \cos^2\theta S_{Z}(0)$.
$\Gamma_{r} = \Gamma_{r}^\mathrm{det} + \Gamma_{r}^\mathrm{env}$,
is the relaxation rate, with $\Gamma_{r}^\mathrm{det}= \nu^2
\cos^2\theta S_{Y}(\phi)$ and $\Gamma_{r}^\mathrm{env}=
\cos^2\theta S_{Z}(\phi)$. We have also defined $\lindblad[A]\rho
= A \rho A^\dag - (A^\dag A \rho + \rho A^\dag A )/2$.

\begin{figure}
\subfigure{\includegraphics[height=2.5cm]{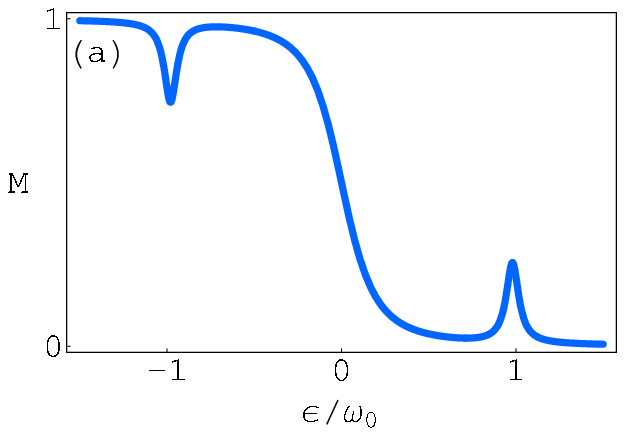}
\label{fig:DCcurrent}}
\subfigure{\includegraphics[height=2.5cm]{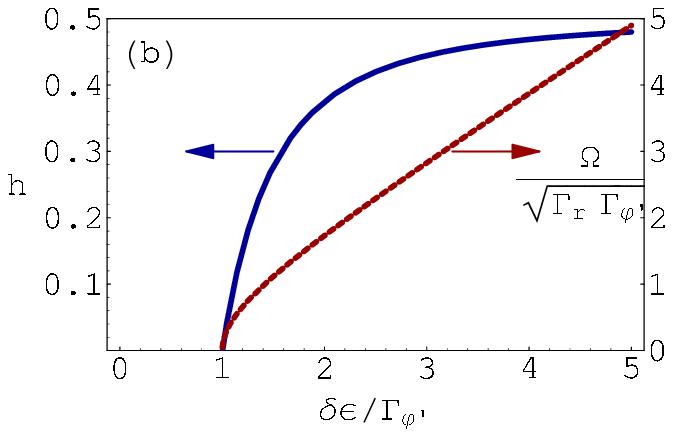}\label{fig:PeakHeightOmegaVsEpsilon}}
\subfigure{\includegraphics[width=8cm]{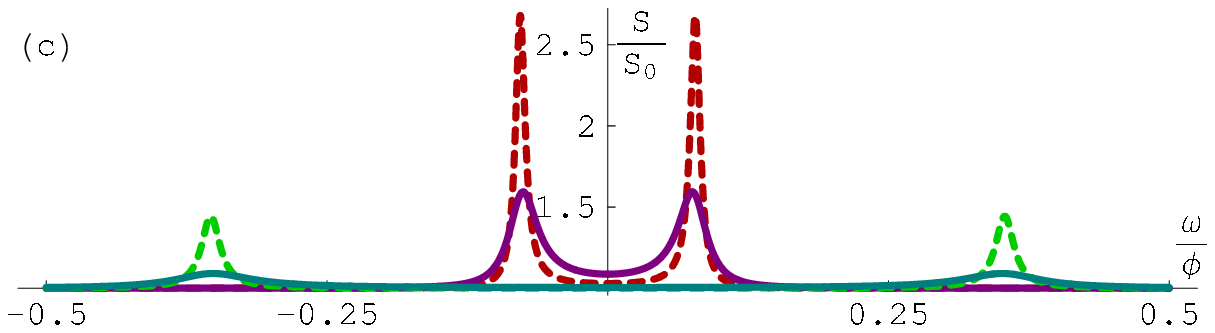} \label{fig:PS}}
\caption{(a) Variation in conductance (scaled between 0 and 1)
versus dot bias.  Parameters
$\Gamma_{\varphi}=\Gamma_r=\Omega=0.03\omega_0$ and
$\Delta=0.2\omega_0$ were taken to be constant over the entire
range of $\epsilon$.  (b) The peak height $h$ as a function of
peak width $\delta\epsilon$.  Increasing microwave power scans
from left to right.  Also shown is the relationship between
$\delta\epsilon$ and the Rabi frequency $\Omega_0$. (c) Power
spectrum, $\tilde S(\omega)$.  Inner peaks: $\theta=\pi/20$, outer
peaks: $\theta=\pi/4$.  The tall peaks have detector limited
dephasing, $\Gamma_\varphi=\Gamma_\phi^\mathrm{det}=2\nu^2 eV
\cos^2{\theta}$, whilst the short peaks are dominated by phonon
dephasing, $\Gamma_\varphi=0.03/\phi$.  Other parameters are
$eV/\phi=0.5$, $\nu=0.1$, $\Omega_0/\phi=0.5$.}
\end{figure}

\paragraph{DC conductance}
The DC current through the QPC is related to the steady state
occupation probability of the dot nearest the QPC, to first order
in $\nu$, by $I=I_0-\delta I \bra{l}\rho(\infty)\ket{l}$ where
$I_0$ corresponds to the current when the electron is localized in
state $\ket{r}$ and $\delta I = 4\nu \calT e V$
\cite{stace:136802,sta03c}. Using \eqn{eqn:mastereqn2}, we can
compute the steady state occupation of the left well, and the
scaled conductance ($M=1+(I-I_0)/\delta I = 1 -
\bra{l}\rho(\infty)\ket{l}$) is given by
\begin{equation}
M=\frac{1}{2} - \frac{\epsilon
\Gamma_r(\eta^2+\Gamma_{\varphi'}^2)}
     {2\phi[\eta^2 \Gamma_r + \Gamma_{\varphi'}(\Omega^2 + \Gamma_r
     \Gamma_{\varphi'})]}, \label{eqn:M}
\end{equation}
where $\Gamma_{\varphi'}=\Gamma_\varphi+\Gamma_r/2$ is the total
dephasing rate. $M$ is plotted as a function of $\epsilon$ in Fig.
\ref{fig:DCcurrent}. Note that Fig. \ref{fig:DCcurrent} is in
excellent qualitative agreement with recent experimental
observations \cite{pet04,YaleDrivenQubit,Duty04}. The absence of
multi-photon resonances in Fig. \ref{fig:DCcurrent} is a
consequence of the first RWA made above.

The resonances occur at $\epsilon=\sqrt{\omega_0^2-\Delta^2}$.
Useful spectroscopic information may be extracted from these
resonant peaks. If the driving frequency, $\omega_0$, is known,
and the Rabi frequency, $\Omega$, is known independently (e.g.
from observations of the time dependence of the detector current,
as discussed below) then $\Gamma_r$ and $\Gamma_{\varphi'}$ can
both be determined. When $\omega_0 \gg \Delta$ and assuming that
$\Gamma_r$, $\Gamma_{\varphi'}$ and $\Omega$ do not vary
significantly across the peak, from \eqn{eqn:M} we find $\Omega^2
\approx{\Gamma_r\Gamma_{\varphi'}({\delta\epsilon^2}/{\Gamma_{\phi'}^2}-1)}$,
where $\delta\epsilon$ is the half-width-half-maximum for the
peak. Therefore, plotting $\Omega$ against $\delta \epsilon$
allows both $\Gamma_r$ and $\Gamma_{\varphi'}$ to be determined.
However, in the absence of time-resolved measurements, $\Omega$
may be unknown, because the relationship between the input
microwave power and the electric field coupling to the qubit may
be unknown. In this case, $\Gamma_{\varphi'}$ can still be
extracted by plotting the peak height, $h$, against $\delta
\epsilon$, for different values of the incident power. Again
assuming $\omega_0 \gg \Delta$ and that $\Gamma_r$,
$\Gamma_{\varphi'}$ and $\Omega$ do not vary significantly across
the peak, $h$ and $\delta \epsilon$ are related by $h \approx
{1}/{2}-{\Gamma_{\varphi'}^2}/{ 2\delta\epsilon^2}$, which is
independent of the (unknown) quantity $\Omega$. These results are
shown in figure \ref{fig:PeakHeightOmegaVsEpsilon}. Note that, for
sufficiently weak driving, the peak width directly gives
$\Gamma_{\varphi'}$, while for stronger driving, the peak width is
proportional to $\Omega$.

\paragraph{Power Spectrum}

Further spectroscopic information may be obtained from the power
spectrum of the current through the QPC, which is given by
$S(\omega)=2 \int_{-\infty}^{\infty} d\tau G(\tau) e^{-i\omega
t}$, where $G(\tau) = E[I(t+\tau)I(t)]-E[I(t+\tau)]E[I(t)]$, is
the current autocorrelation function, and $E[\ldots]$ denotes the
classical expectation. $S(\omega)$ can be computed using the same
formalism as in \cite{Goan2001b,sta03c}. In the low-bias regime
the scaled (symmetric) power spectrum $\tilde
S(\omega)={S(\omega)}/{S_0}$ (where $S_0=2e I_{DC}=e^2\calT^2 eV$
is the shot noise background) is closely approximated by
\begin{equation}
\tilde
S(\omega)=\frac{s_{\Omega'}\gamma_{\Omega'}^2}{\gamma_{\Omega'}^2+(\omega-{\Omega'})^2}+\frac{s_{\Omega'}\gamma_{\Omega'}^2}{\gamma_{\Omega'}^2+(\omega+{\Omega'})^2}+
\frac{s_0 \gamma_0^2}{\gamma_0^2+\omega^2}+1,
\end{equation}
where
\begin{eqnarray}
\gamma_{\Omega'}&=& \frac{2\left( 2{\eta }^2 + {\Omega }^2 \right)
{{\Gamma }_{\varphi}} +
    \left( 2{\eta }^2 + 3{\Omega }^2 \right)
     {{{\Gamma }}_r}}{4\left( {\eta }^2 + {\Omega }^2 \right) }
      , \nonumber\\
\gamma_0&=&\frac{2{\Omega }^2{{\Gamma }_{\varphi}} +
    \left( 2{\eta }^2 + {\Omega }^2 \right)
     {{{\Gamma }}_r}}{2\left( {\eta }^2 + {\Omega }^2 \right) }
      ,\nonumber\\
s_{\Omega'} &=& \frac{4{\Omega }^2{{\Gamma
}_{\varphi}^\mathrm{det}}}
  {  2( 2{\eta }^2 + {\Omega }^2) {{\Gamma }_{\varphi }}+( 2{\eta }^2 + 3{\Omega }^2 )
     {{{\Gamma }}_r}},\nonumber\\
s_0 &=&\frac{4{\eta }^2{\Omega }^2{{\Gamma
}_{\varphi}^\mathrm{det}}
    ( {\Omega }^2{{{\Gamma }_r}}^2 +
      4( 2{\eta }^2 + {\Omega }^2 ) {{\Gamma }_r}
       {{\Gamma }_{\varphi }} + 4{\Omega }^2{{{\Gamma }_{\varphi }}}^2 ) }
    {{(     2{\Omega }^2{{\Gamma }_{\varphi }} +( 2{\eta }^2 + {\Omega }^2 ) {{\Gamma }_r} ) }^3}
     .\nonumber
\end{eqnarray}
Thus the positions of the peaks yield the Rabi frequency
$\Omega'$, while the widths of the peaks contain other important
spectroscopic information about the qubit. The height of the peaks
at $\pm \Omega'$ is maximised when the external field is resonant
with the qubit transition, $\eta=0$. Note that $\tilde{S}(\Omega')
\le 2$ and thus the peak heights are no more than three times the
shot noise background, as shown in Fig. \ref{fig:PS}. The bound
$\tilde{S}(\Omega') = 2$ can only be met when dephasing is
detector dominated, since extra dephasing due to the environment
reduces the height of the peaks.

\begin{figure}[t]
\subfigure{\includegraphics[height=2.5cm]{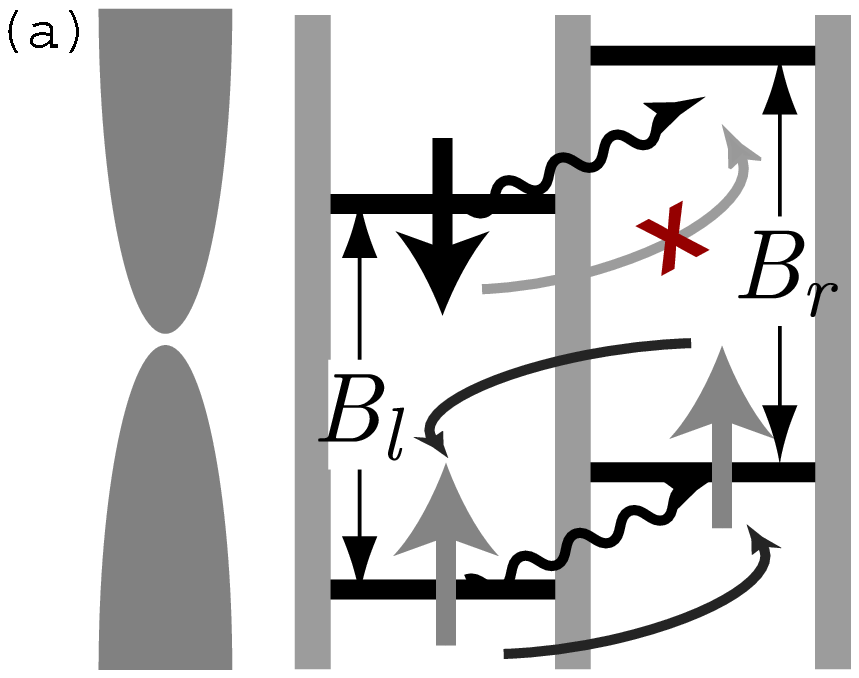}\label{fig:twospincartoon}}\hfil
\subfigure{\includegraphics[height=2.5cm]{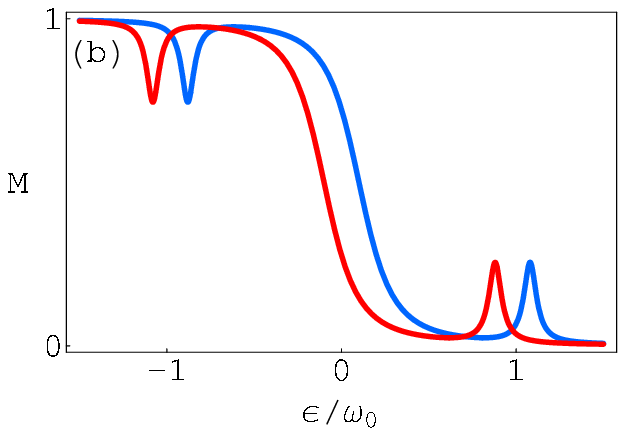}\label{fig:spinDC}}
\caption{(a) Schematic of spin measurement scheme. (b) Conductance
curves for spin-up and spin-down configurations showing distinct
resonance peaks.}
\end{figure}

\paragraph{Spin measurement}
Motivated by the preceding analysis, we propose a method for
single shot spin readout, using a microwave driving field and an
inhomogeneous Zeeman splitting across the CQD. This splitting
could be generated by an inhomogeneous magnetic field, or
engineering different $g$-factors in each dot. In this case, one
spin configuration, say spin-down, may be made resonant with the
driving field, whilst the other, spin-up, is detuned by an amount
$\eta$, and thus the spin can be determined by observing the
current through the QPC [Fig. \ref{fig:twospincartoon}]. Such a
scheme is analogous to the method of spin readout via quantum
jumps in atomic systems \cite{Blatt1988}. We note that alternative
methods for spin readout have also been proposed, via inhomogenous
fields but without driving \cite{Engel2004}, and by driving
spin-flip transitions within a single dot \cite{Friesen2004}.

Spin can be included in the Hamiltonian by replacing $\sigma_x$
and $\sigma_z$ in Eqs. (\ref{eqn:hsys}-\ref{eqn:hphon}) with
generalized tunnelling and bias operators, i.e. $\sigma_x \to c_{l
\uparrow}^{\dagger}c_{r \uparrow}^{\phantom{\dagger}}+c_{l
\downarrow}^{\dagger}c_{r \downarrow}^{\phantom{\dagger}} + \hc$
and $\sigma_z \to n_{l}-n_{r}$ where $n_i = c^\dag_{i
\uparrow}c_{i \uparrow}+ c^\dag_{i \downarrow}c_{i \downarrow}$
denotes the number of electrons on site $i$. The inhomogenous
Zeeman splitting is included by adding the term
\begin{equation}
H_{\mathrm{Zeeman}} = - \frac{1}{2} \sum_{i=l,r} B_i S^{(z)}_{i}
\end{equation}
to $\hsys$, where $B_i$ denotes the Zeeman splitting on each site,
and $S^{(z)}_{i}=  c^\dag_{i \uparrow}c_{i \uparrow} - c^\dag_{i
\downarrow}c_{i \downarrow}$. We neglect any explicit coupling
between spin up and spin down states (processes which can induce
such transitions can be accounted for by a finite spin flip time,
which we discuss below).

The Zeeman term amounts to a spin dependent bias between the dots,
$\epsilon_{\downarrow / \uparrow} = \epsilon \pm (B_r-B_l)/2$. The
DC response of the detector is therefore shifted for each spin
configuration, as shown in Fig. \ref{fig:spinDC}. If $B_r-B_l
\gtrsim \delta \epsilon$, the resonant peaks are clearly resolved.
In this case, if $\epsilon$ is tuned such that the spin-down
transition is resonant with the driving field, the currents for
each spin configuration are approximately $I_\uparrow \approx I_0
- \delta I$ and $I_\downarrow \approx I_0 - \delta I / 2$. If the
detector is shot noise limited, the time taken to resolve these
currents is $\tau^{-1}_{\uparrow\downarrow} =
(I_\uparrow^{1/2}-I_\downarrow^{1/2})^2/2e \approx \delta I^2/32
I_0$. Thus $\tau_{\uparrow\downarrow} = 4 \tau_{01}$ where
$\tau_{01}$ is the time taken to distinguish the currents due to 0
and 1 electrons on the dot adjacent to the QPC. It is believed
that shot noise limited QPC detectors will be available in the
near future with $\tau_{01} \sim 25$ ns \cite{van04}, and so
$\tau_{\uparrow\downarrow} \sim 100$ ns should be possible. This
compares favourably with recently observed spin flip times ($T_1
\approx 1$ ms) in GaAs quantum dots \cite{Elzerman2004}.

The use of microwave driving for spin readout offers some
advantages over other schemes where no driving is used. Firstly, a
relatively small differential Zeeman splitting is needed. In order
to clearly resolve the resonant peaks, we require $B_r-B_l \gtrsim
\delta \epsilon$. The lower bound on the peak width is given by
$\delta \epsilon >  \Gamma_{\varphi'}$. So for a charge dephasing
rate of $\Gamma_{\varphi'} \sim 10^8$ s$^{-1}$ \cite{hay03}, we
require $B_r-B_l \gtrsim 0.07$ $\mu$eV. If one attempts readout
via an inhomogenous Zeeman splitting but without driving
\cite{Engel2004}, then to obtain a comparable signal-to-noise, the
differential Zeeman splitting must be larger than the central
transition region in Fig. \ref{fig:spinDC}, i.e. $B_r-B_l \gtrsim
\max (\Delta, k T)$. For $T = 100$ mK, $B_r-B_l \gtrsim 9$ $\mu$eV
is required. Thus in GaAs ($g=0.44$), with a uniform field of 1 T,
our scheme requires a $g$-factor variation between the dots of
$\Delta g / g\sim$ 0.3 \%, whereas without driving, one would
require $\Delta g / g \sim$ 35 \%.

Secondly, the scheme can be used in such a way that a definite
signal is always obtained for both spin configurations. Switching
the microwave frequency first on resonance with the spin-down
transition, and then on resonance with the spin-up transition
yields a definite signal indicating the spin state. That is, when
the driving frequency is switched, the DC conductance will
increase if the electron is spin-up, and will decrease if it is
spin-down. This is in contrast to other measurement schemes in
which a definite signal is only registered for one spin
configuration, \cite{Engel2004,van04,Barrett04}, with the other
state indicated only by the lack of a signal. This has the
advantage of eliminating false negative signals, where an
``no-signal'' event is erroneously recorded as evidence for a
particular spin configuration. Thus the resulting measurement
fidelity should be improved.

In summary, we have analysed the dynamics of a continuously
observed, driven solid state qubit, coupled to a generic bosonic
environment. Both the environment and the coupling to the detector
contribute to the qubit relaxation and dephasing rates. Useful
spectroscopic information, in particular the dephasing rate
$\Gamma_{\varphi'}$, can be extracted from DC measurements of the
detector output current alone, even when the coupling between the
microwave field and the qubit is unknown. If the power spectrum of
the detector output noise can also be measured, then the
relaxation rate $\Gamma_{r}$ and Rabi frequency $\Omega$ can also
be determined. We have also proposed a single shot spin readout
technique using microwave driving, which offers advantages over
existing schemes and can be implemented with current technology.

We thank Gerard Milburn, Charles Smith, Andrew Doherty, Jason
Petta, Charlie Marcus, Bill Munro and Tim Spiller for useful
conversations. SDB thanks the EU NANOMAGIQC project
(IST-2001-33186) for support. TMS is funded by the CMI-Fujitsu
collaboration.

\bibliographystyle{apsrevnourl}
\bibliography{mwpaperFolded}

\end{document}